\documentclass[12pt,a4paper]{article}

\usepackage{psfig,latexsym}

\newcommand{\e}{\mbox{e}}
\newcommand{\ri}{\mbox{$R_{\mbox{\scriptsize inc}}$}}

\newcommand{\ebind}{\mbox{$E_{\mbox{\scriptsize bind}}$}}
\newcommand{\kb}{\mbox{$k_{\mbox{\scriptsize B}}$}}
\newcommand{\ev}{\mbox{eV}}

\newcommand{\plot}[1]{
  \centerline{ \psfig{file=#1,width=0.8\textwidth} }
}

\renewcommand{\title}[1]{
  \begin{center}
    \Large \bf #1
  \end{center}
}

\renewcommand{\author}[1]{
  \centerline{\sc \large #1}
}

\renewcommand{\thanks}[1]{
  \footnote{#1}
}

\renewcommand{\abstract}[1]{
  \begin{quote} #1  \end{quote}
}

\newcommand{\institute}[1]{
  \begin{center}
    \em #1
  \end{center}
}

\newcommand{\PACS}[1]{   
{\sc Pacs:} \parbox[t]{12cm}{\em #1}
}

\newenvironment{acknowledgement}[0]{
  \begin{center} \bf *** 
  \end{center}}{}

\begin{document}

\title{Ehrlich--Schwoebel barrier controlled slope selection in
  epitaxial growth}
\author{S. Schinzer, S. K\"ohler, G. Reents}
\institute{Universit\"at W\"urzburg, Institut
         f\"ur Theoretische Physik, Am Hubland D-97074 W\"urzburg,
         Germany}
\abstract{
%\begin{abstract}
We examine the step dynamics in a 1+1 dimensional model of epitaxial
growth based on the BCF--theory. The model takes analytically into
account the diffusion of adatoms, an incorporation mechanism and an
Ehrlich--Schwoebel barrier at step edges. We find that the formation
of mounds with a stable slope is closely related to the presence of an
incorporation mechanism. We confirm this finding using a
Solid--On--Solid model in 2+1 dimensions. In the case of an
infinite step edge barrier we are able to calculate the saturation
profile analytically. Without incorporation but with inclusion of
desorption and detachment we find a critical flux for instable
growth but no slope selection. In particular, we show that
the temperature dependence of the selected slope is solely
determined by the Ehrlich--Schwoebel barrier which opens a new
possibility in order to measure this fundamental barrier in
experiments. 
%\end{abstract}
}


\section{Introduction}
Molecular beam epitaxy (MBE) has attracted much interest from both,
theoretical and experimental physicists. On the one hand it allows the
fabrication of high quality crystals with arbitrary composition and
modulated structures with atomically controlled thickness
\cite{hs96}. On the other hand it represents a model of nonequilibrium
physics which still lacks a general theory \cite{bar95}. In
particular, the appearance and the dynamics of three dimensional (3D)
structures (pyramids or mounds) in crystal growth are not well
understood in terms of the underlying microscopic processes.

%Several theoretical descriptions of growth exist \cite{bar95}. Rate
%equations \cite{cpp89}, lattice models (in particular the
%solid--on--solid models) \cite{sv93,tks98,sp96}, or continuum
%equations \cite{sp94} have been applied to study crystal growth. Such
%diverse issues as the island size scaling in submonolayer epitaxy
%\cite{bru98}, modeling of RHEED--oscillations \cite{sv93,hfa97}, or
%the investigation of scaling behaviour in thin film growth
%\cite{kru97b} have been addressed.

A long time ago Burton, Cabrera and Frank introduced the BCF--theory
of crystal growth \cite{bcf51}. Within this theoretical approach the
crystal surface is described by steps of single monolayer height.
The evolution of the surface is calculated by solving
the diffusion equation on each terrace. Within this framework the
growth of spirals and the step flow has been investigated. Elkinani
and Villain investigated such a model including the nucleation
probability of new islands \cite{ev94}. They found that the resulting
structures are unstable. Towers appear which keep their lateral
extension and grow in height only. They called this effect the
Zeno--effect. The same observation has been made with a ``minimal
model'' of MBE where fast diffusion together with a high
Ehrlich--Schwoebel barrier has been implemented \cite{kru97a}.

Even though the Zeno--effect has been observed recently on Pt(111)
\cite{ksc99}, quite typically a coarsening process with appearance of
slope selection emerges which has been reported for such diverse
systems as
% Fe(001), Cu(001), GaAs(001), or HgTe(001) \cite{sps95,oes96}. 
Fe(001) \cite{tkw95,sps95}, Cu(001) \cite{eff94,zw97}, GaAs(001)
\cite{joh94,ojl95}, and HgTe(001) \cite{oeg98}.  In addition, slope
selection seems to be the generic case of solid--on--solid computer
simulations \cite{sv95,ojl95,tks98}.

In terms of continuum equations the selection of a stable slope has
been related to the compensation of uphill and downhill currents
\cite{sp94,sp96}. An uphill current can be generated by an
Ehrlich--Schwoebel barrier \cite{eh66,ss66}. The
barrier hinders adatoms to jump down a step edge. Hence, more
particles attach to the upper step edge which leads to a 
growth--instability \cite{vil91} and 3D--growth.

Another process, which constitutes a downhill current, has been
recognized using molecular dynamics simulations
\cite{est90,yhp98}. Such diverse mechanisms as downward funneling,
transient diffusion or a  knockout process at step edges lead to the
incorporation of {\em arriving} particles at the lower side of the
step edge. In addition, it has been suggested that such a process is
responsible for reentrant layer--by--layer growth \cite{swv93-1}

Recently we have proposed a simplified model of epitaxial growth
quite similar to the ``minimal model'' of Krug \cite{bks98}. In
particular we found that an incorporation mechanism is crucial to
achieve slope selection. However, one simplifying assumption of the
model is an infinite Ehrlich--Schwoebel barrier. 

In this article we will present in more detail the argument leading to
slope selection and we will generalize our results using a continuous
step dynamics model analogous to \cite{ev94}.  In sec.~\ref{BCF} 
we will introduce our extension of the BCF theory and will
discuss the relation to existing results (sec.~\ref{BCF} and
\ref{closure}). Typical mound morphologies 
and the growth dynamics are compared in section \ref{growth}.
Afterwards we will investigate the emergence of slope selection within
the framework of this model (sec. \ref{slopeselection}). We will show
that the selected slope has a temperature--dependence which is solely
determined by the Ehrlich--Schwoebel barrier. Hence, the determination
of selected terrace widths in experiments would give direct insight
into microscopic properties such as the Ehrlich--Schwoebel
barrier. We confirm the predicted importance of the incorporation
mechanism using a kinetic Monte--Carlo simulation of a
Solid--On--Solid model in sec.~\ref{sos-incop}.
Another effective downward current could be due
to detachment from steps and subsequent desorption. We will show in
sec.~\ref{detachment} that slope selection cannot be achieved by these
two processes alone. In section 
\ref{saturation} we will calculate the saturation profile in the
limiting case of an infinite Ehrlich--Schwoebel barrier.

%-----------------------------------
\section{BCF theory}     \label{BCF}

\begin{figure}
\plot{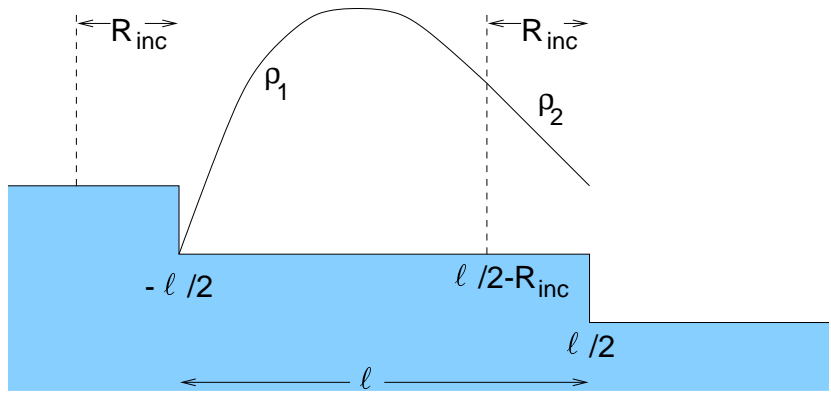}
\caption{\label{crystal}
Density of diffusing adatoms on a
terrace of size $\ell$. The origin of the x-axis is chosen to be in
the middle of the terrace.
}
\end{figure}

The model is based on the Burton--Cabrera--Frank model in 1+1
dimensions. Within this framework the crystal surface is specified by
the position and direction (upward or downward) of steps. Figure
\ref{crystal} shows the crystal surface from the point of view of the
BCF--theory. It is a coarse grained view -- the detailed positions of
atoms are not important. However, the terraces of the height of one
atomic monolayer (ML) can still be distinguished. The most fundamental
assumption is that at each time $t$ the adatom concentration $\rho$ is
a function of the step positions only. In other words, the diffusion
of adatoms is considerably faster than the step velocity. Thus, the
diffusion equation becomes
\begin{equation} \label{diffusion}
\frac{\partial \rho}{\partial t}(x,t) = 0 
                   = D \nabla^2 \rho (x,t) + \frac{F}{a} 
\end{equation}
where $D$ is the diffusion constant and $F/a$ is the flux density with
$a$ denoting the lattice constant. Hence, $1/F$ is the time necessary
in order to deposit one monolayer. Up to now, this equation was
solved with special boundary conditions at $x=-\ell/2$ and
$+\ell/2$ in the literature. Clearly, the boundary conditions are
chosen depending on 
whether the terrace is a vicinal, a top, or a bottom terrace. In the
following we will discuss the typical case of a vicinal terrace. The
extension to top and bottom terraces is straightforward.

To include an incorporation mechanism it is necessary to extend the
theory. We assume that there exists an incorporation radius such that
all particles arriving close to a downward step within this radius
immediately jump down the step edge. Hence, one has to split the
density of diffusing particles into two regions. The first region
close to the upper edge where {\it eq.}~(\ref{diffusion}) holds, and
the second one given by the incorporation radius close to the downward step
where no particles arrive ($F=0$). To describe the motion of steps the
flux of incorporated particles must be taken into account separately.

In the following we will discuss in detail the situation $\ell > \ri$
as sketched in fig.~\ref{crystal}. For smaller terraces only one
region exist and the calculations are much easier. Since 
our analytical calculations will show that $\ell > \ri$ is the generic
case we concentrate on this situation.

The general one-dimensional solution of {\it eq.}~(\ref{diffusion})
is a parabola characterized by three parameters. In addition to the
two diffusion equations, four 
boundary conditions are necessary to determine the two distributions
$\rho_1$ and $\rho_2$ .
\begin{eqnarray}
\rho_1 (-\ell/2) & = & 0  \label{Bup} \\
\rho_1 (\ell/2-\ri) & = & \rho_2 (\ell/2-\ri) \label{Br} \\
\rho_1' (\ell/2-\ri) & = & \rho_2' (\ell/2-\ri) \label{Bj} \\
-D \rho_2'(\ell/2) & = & \frac{D}{\ell_1} \rho_2 (\ell/2) \label{Bdown}
\end{eqnarray}
Condition (\ref{Bup}) is for the special case of perfect absorbing
step edges. (\ref{Br}) and (\ref{Bj}) are necessary to obtain a smooth
density between region 1 and 2. The left hand side of (\ref{Bdown}) is
the particle current at the step edge. On the right hand side this is
reformulated using the number of jump attempts $D \rho_2 (\ell/2)$
multiplied by the probability of overcoming the Ehrlich--Schwoebel
barrier $E_S$. This probability is expressed as the inverse of a typical
length $\ell_1$
\begin{equation}
 \frac{1}{\ell_1} = \frac{1}{a}
               \exp\left(-\frac{E_S}{\kb T} \right)
\end{equation}
where $a$ stands for the lattice constant.

The resulting density distribution has the form as indicated in
fig.~\ref{crystal}: a parabola in the upper region and linear close
to the downward step. The detailed expressions of $\rho_1$ and $\rho_2$
are not of much interest since the evolution of the crystal is determined by
the currents at the edges. In the following we will call $u(\ell)$ the
upward current, {\it i.e.} 
\begin{eqnarray}
u(\ell) & = & -D \rho_1'(-\ell/2)  \nonumber\\
        & = & - \frac{F}{2a\,(\ell + \ell_1)}
          \left( \ell^2 + 2 \ell \ell_1 - 2 \ri \ell_1 - \ri^2 \right).
 \label{jup}
\end{eqnarray}
The downward current due to diffusion (the contribution of the
incorporation mechanism is not included) becomes
\begin{eqnarray}
d(\ell)  & = & -D \rho_2'(+\ell/2)  \nonumber\\
         & = & \frac{F}{2a\,(\ell + \ell_1)} \left( \ell - \ri \right)^2.
 \label{jdown}
\end{eqnarray}
Note, that these results are very similar to the corresponding
equations (2.2) and (2.3) of reference \cite{ev94} where no
incorporation was considered. Setting $\ri=0$ we regain their
results. 

The absence of a dependence on $D$ reflects the ansatz of a
quasi--stationary distribution. All arriving particles are compensated
for by the loss of particles at the borders and hence the currents are
proportional to $F$. The density itself is proportional to
the ratio $F/D$ which again is intuitively clear.

\begin{figure}
\plot{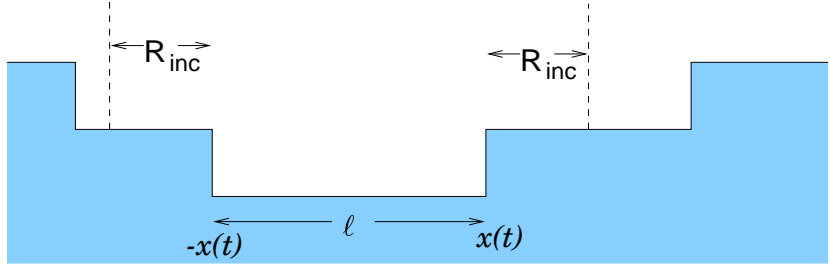}
\caption{ \label{bottom}
Arrangement of steps around the bottom terrace of width $\ell$ between
two mounds. 
} \end{figure}

\section{Closure of bottom terraces}                   \label{closure}
In the following, we will reinvestigate the discussion of \cite{ev94}
concerning the closure of a bottom terrace 
({\it c.f.}~fig.~\ref{bottom}). In the limiting case of an infinite
Ehrlich--Schwoebel barrier the dynamics of the steps become very
simple. We denote by $x(t)$ the position of the right step which of
course will depend on the time $t$. The origin is chosen to be in the
middle of the bottom 
terrace. Due to the infinite Ehrlich--Schwoebel barrier the movement
of the right and the left step are symmetric. The evolution is then
described by  
$\dot{x}(t) = -F  x(t) - F  \ri$.
The first term corresponds to the particles which do fall on the
bottom terrace and diffuse to the right. The second term is the
contribution of particles which are incorporated from the step above
(which is valid as long as the bottom terrace is more than a distance
\ri\ away from a top terrace). As a result $x(t)$ evolves as
\begin{equation} \label{xbottom}
x(t) = (x_0 + \ri) \exp\left(-F t\right) - \ri.
\end{equation}
As long as $\ri > 0$ there exists a closure time
\begin{equation} \label{tc}
t_c = \frac{1}{F} \ln \left( \frac{x_0 + \ri}{\ri} \right).
\end{equation}
Without an incorporation mechanism (\ri=0) the bottom terrace will
never be closed. This is the reason why Elkinani and Villain called
their model the Zeno--model to remind the greek philosopher and his
paradox. Even though the situation is changed if the discrete
structure of the terraces is considered\footnote{The currents can be
  translated into probabilities of placing a particle at the step
  edge. Hence, a bottom terrace of width one always has a nonvanishing
  probability to be filled.} they showed that this trend still holds
which gives rise to the formation of deep cracks. Likewise they
found that even finite values of the Ehrlich--Schwoebel barrier do
not change this growth scenario which has been investigated in more
detail in \cite{pol97}. Once mounds are built up they remain
forever with a fixed lateral size. Our discussion of this limiting
case shows that the inclusion of an incorporation mechanism changes
the growth in a fundamental manner.

\section{Growth dynamics}  \label{growth}
\begin{figure}[t]
\setlength{\unitlength}{0.0941\textwidth}      % {0.518cm}
\begin{picture}(8.5,3)            % 16.4,5.3)
\put(0,0){\parbox[b]{4.3cm}{\psfig{file=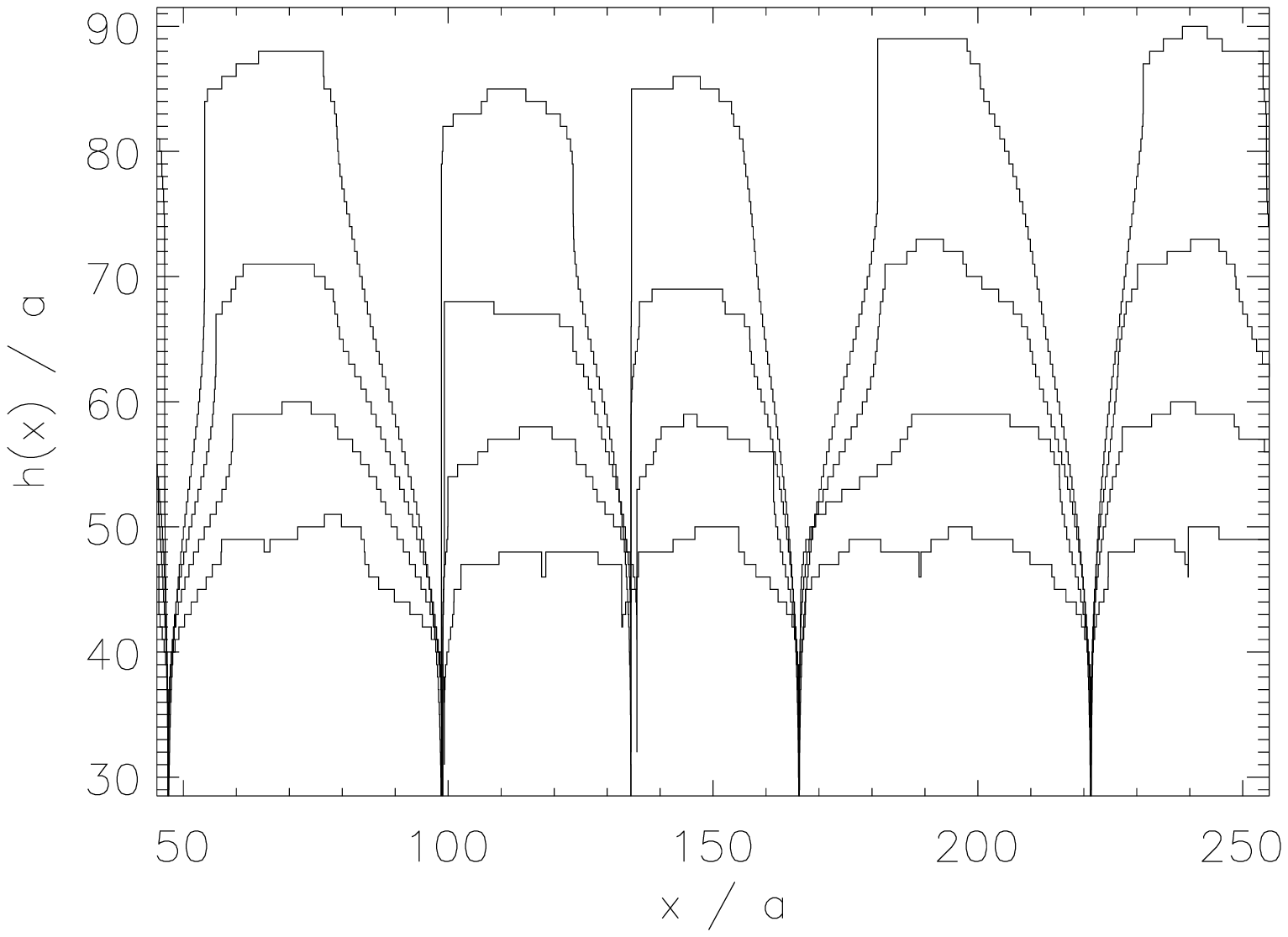,width=0.38\textwidth}}}
\put(4.25,0){\parbox[b]{4.3cm}{\psfig{file=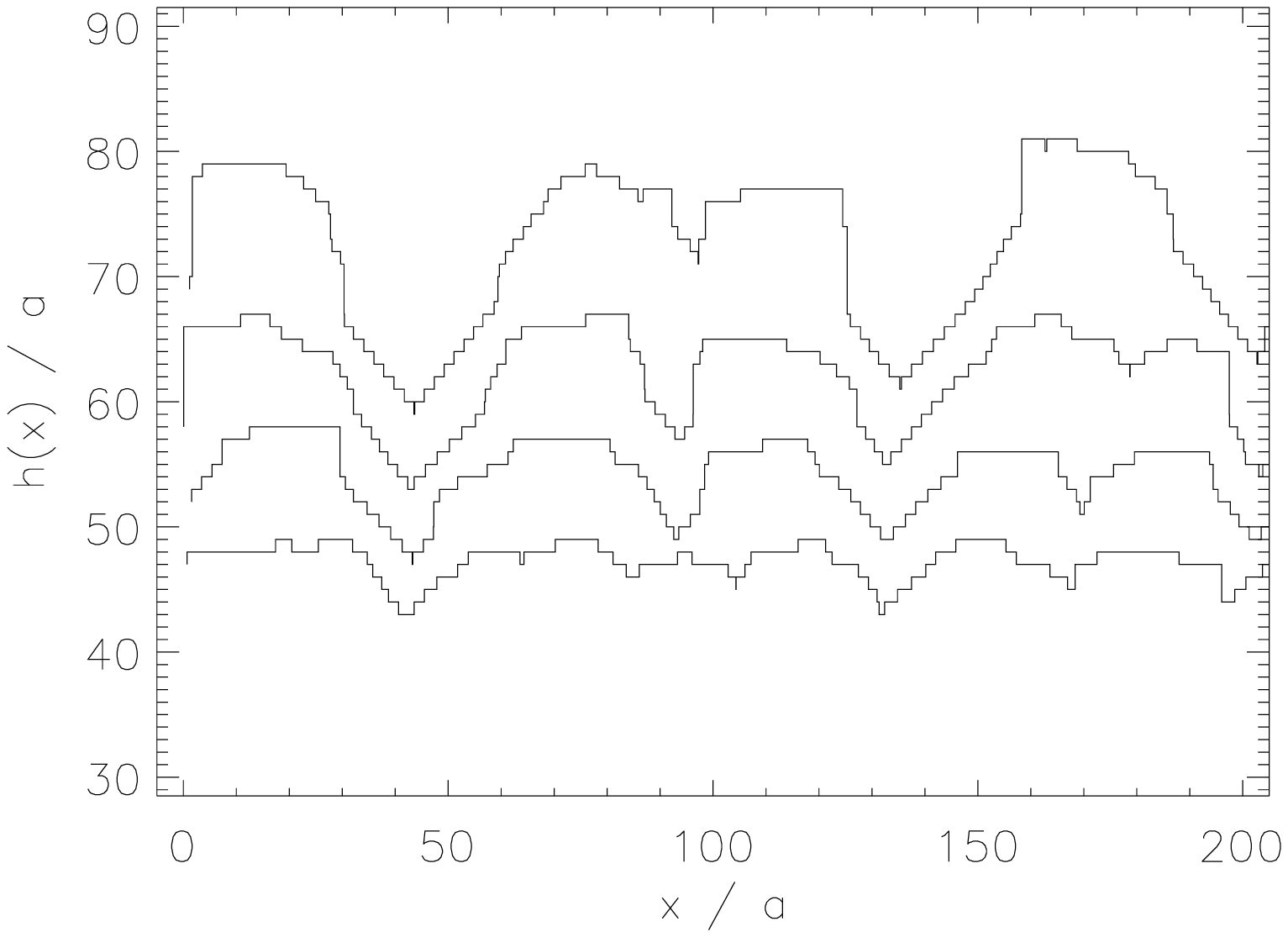,width=0.38\textwidth}}}
\put(0.93,0.7){\bf a)}
\put(5.15,0.7){\bf b)}
\end{picture}
\caption{ \label{profiles} 
  In (a) we show the typical height profile of the Zeno model. With
  inclusion of an incorporation mechanism in (b) we obtain structures with
  slope selection and prevent the formation of deep cracks. In both
  cases we have deposited (from bottom to top) 20, 50, 100, and 200 ML
  where the height profiles have been shifted by +25, 0, -45, and -140
  ML respectively. The regions are subsections of $200\,a$
  out of a surface of size $485\,a$.
}
\end{figure}

To set up the basic ideas of the behaviour during crystal growth we
show two typical surface profiles according to the numerical
integration of the step system. In fig.~\ref{profiles} we compare
the resulting structure of the Zeno model \cite{ev94} without an
incorporation mechanism and with the inclusion of such a mechanism. 

The simulations were carried out on on a system of $485\,a$ width
with parameters corresponding to the model used in sec.~\ref{sos-incop}.
\begin{eqnarray*}
D & = & 10^{12}\, \exp\left( -\frac{0.9\, \ev}{550 \mbox{K} \, \kb} \right)
\frac{a^2}{\mbox{s}}  \approx 5664\, \frac{a^2}{\mbox{s}} \\
\ell_1 & = & 
       \exp \left( +\frac{0.1\, \ev}{550 \mbox{K} \, \kb} \right)\, a
       \approx 8.2\, a \\
\ri & = & 1\, a \\
F & = & \mbox{1 ML s}^{-1}
\end{eqnarray*}
As in \cite{ev94} the Ehrlich--Schwoebel barrier has been suppressed
for bottom terraces of one lattice constant width.
Without an additional incorporation mechanism the
appearance of trenches is unavoidable in accordance to
\cite{pol97}. The incorporation mechanism gives rise to a well defined
slope which does not change with time.
Another fundamental difference is the coarsening 
behaviour. Without an incorporation mechanism the trenches are stable
and the number of mounds remains constant. The additional
incorporation mechanism leads to a coarsening behaviour. 

In lattice models as well as for continuum equations the coarsening is
driven by fluctuations \cite{tsv98,bks98} and in 1+1 dimensions the
corresponding exponent is 1/3. This is in accordance to
Ostwald-ripening which has been predicted from the similarities of the
relevant continuum equations \cite{sp96}. However, since we treat the
step evolution in a deterministic manner we do not obtain a scaling
behaviour. The only way fluctuations come into play during the
simulation is when new islands are nucleated. As a consequence the
evolution of {\it e.g.}~the width of the height distribution $w$ is
characterized by jumps (data not shown). A jump in $w$ appears each
time when two mounds merge.  These findings are a direct confirmation
of the relevance of the fluctuations for the coarsening behaviour.

%-----------------------------------------------
\section{Slope selection}  \label{slopeselection}
The inclusion of an incorporation mechanism leads to slope selection
which is apparent from fig.~\ref{profiles}b. Siegert and Plischke
\cite{sp94} required a cancelation of upward and downward currents in
the continuum equations. Again, in the case of an infinite
Ehrlich--Schwoebel barrier the calculations are straightforward.
 In this case the
downward current on a vicinal terrace of size $\ell$ is 
solely due to the incorporation mechanism, {\it i.e.} proportional to
$F\ri$. All the remaining diffusing adatoms will contribute to the
upward current and hence the current will be $F(\ell-\ri)$. 
As a consequence the slope selection will be achieved with a mean
terrace width of size 
\begin{equation}
  \ell^* = 2 \ri
\end{equation}
in accordance to the findings in \cite{bks98}.

It remains to calculate the terrace widths for arbitrary parameters.
Since we know the currents (equations
(\ref{jup}), (\ref{jdown})  and the incorporation mechanism) we obtain
the overall slope (resp. terrace width) dependent current 
\begin{eqnarray}
J(\ell) & = & u(\ell)+d(\ell) + F \ri \nonumber\\
        & = & \frac{F}{2(\ell + \ell_1)} \left( 2 \ri^2 + 4\ri \ell_1
        - 2 \ell \ell_1 \right)   \label{Jvell}
\end{eqnarray}
Note that a positive $J(\ell)$ signifies a downward current (to the
right in fig.~\ref{crystal}). The stable slope, where no net upward
or downward current remains, is given by the condition $J(\ell)|_{\ell
  = \ell^*}=0$  and yields 
\begin{eqnarray}
  \ell^* & = & 2\ri + \frac{\ri^2}{\ell_1} \nonumber \\
         & = & 2\ri + \frac{\ri^2}{a}  \e^{-E_S/\kb T} \label{ell*}
\end{eqnarray}
As can be seen from expression (\ref{Jvell}) the current is positive
for small values of $\ell$ and becomes negative for $\ell >
\ell^*$. Hence $\ell^*$ is stabilized by the current.

The stable slope does not depend on the diffusion constant. However,
it should be clear from the derivation, that in order to achieve slope
selection the typical diffusion length should be much larger than
$\ell^*$. Otherwise the vicinal terraces would not proceed via step
flow. Rather new nucleation events on the terraces would lead to a
rugged surface structure.

%--------------------------------------------------------
\section{Solid--On--Solid model}        \label{sos-incop}
In order to verify the predicted importance of the incorporation
mechanism we use computer simulations of the Solid--On--Solid (SOS)
model on a simple cubic lattice. All processes on the surface are
(Arrhenius-) activated processes 
which are described by one {\em common} prefactor $\nu_0$ and an
activation energy which is parameterized as follows: $E_B$ is the
barrier for surface diffusion, at step edges an Ehrlich--Schwoebel
barrier $E_S$ is added. However, this barrier is not added for a
particle which sits on top of a single particle or a row
\cite{pkn91,swv93-1}. Each next neighbour contributes $E_N$ to the
activation energy. Within this framework the diffusion
constant becomes $D=\nu_0 \cdot \exp \left( -E_B/\kb T
\right)$. 

\begin{figure}[t]
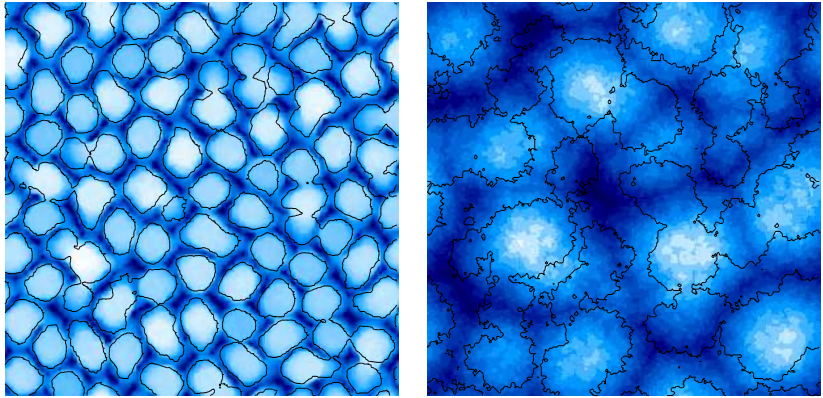

  \setlength{\unitlength}{0.0941\textwidth} 
  \begin{center}
    \begin{picture}(8.5,4.25)
      \put(0,0){\parbox[b]{4.2cm}{\psfig{file=surf22.ceps,width=0.38\textwidth}}}
      \put(4.35,0){\parbox[b]{4.2cm}{\psfig{file=surf42.ceps,width=0.38\textwidth}}}
    \end{picture}
  \end{center}
  \caption{ \label{sos} 
    We compare the morphology of the surfaces without (left) and with
    (right) an incorporation mechanism. Note that the two grey--scales
    are different. The heights in the left picture range from 1208 to
    1326 whereas the right surface only spans a height difference of
    12 from minimum to maximum. The contour lines are drawn for the
    same surface at an earlier stage where only 300 ML have been
    deposited. Without incorporation the mounds (towers) are nearly
    unchanged despite the deposition of 1000 ML.
  }
\end{figure}

Here, we concentrate on a particular set of parameters even though
other parameter sets were used as well.  We choose $\nu_0=10^{12}
s^{-1}$, $E_B=0.9\ev$, $E_N=0.25\ev$, and $E_S=0.1\ev$. This model was
already investigated in \cite{sk98} and reproduces some kinetic
features of CdTe(001). The deposition of particles occurs with a rate
$F$. The two simulations shown in fig.~\ref{sos} are carried out on
a $300 \times 300$ lattice at 560 K and started on a singular (flat)
surface.

In fig.~\ref{sos} the resulting surfaces with and without the
inclusion of the incorporation mechanism are shown. Without an
incorporation mechanism no slope selection occurs. Clearly, without
incorporation the configuration of the towers remains unchanged
whereas the inclusion leads to coarsening. The number of mounds
diminishes with time. Hence, without an incorporation mechanism no
coarsening can be identified. We want to mention that it seems that at
higher temperatures the attachment/detachment kinetics of atoms at
step edges yields a coarsening effect (data not shown). However, still
no slope selection has been observed.

At first glance our findings contradict previous results obtained with
a very similar model. \v{S}milauer and Vvedensky obtained a formation
of mounds with slope selection irrespective of the inclusion or
exclusion of an incorporation mechanism \cite{sv95}. However, they
implemented the Ehrlich--Schwoebel barrier in a different way. Rather
than to hinder the jump over a step edge they impede the jump towards
a step edge. Their motivation for this implementation was to allow the
adatoms to leave a small line of particles of width one which has been
tested as a cause for reentrant layer-by-layer growth
\cite{pkn91,swv93-1}. In our 
simulations the same goal is achieved by suppressing the
Ehrlich--Schwoebel barrier in such a situation. However, in their
simulations particles arriving directly at a step edge have a
probability of 1/4 to jump down the edge, 1/4 to jump away from the
edge and 1/2 to jump along the step edge. Effectively this leads to an
incorporation radius of length 1/2. 

Other simulations of SOS-models used bcc(001) \cite{af96a,tks98} in
order to study the growth of typical metals.  In these simulations the
SOS--restriction is implemented in such a way that an adatom must be
supported by the four underlying atoms. Hence, the downward funneling
process is directly implemented. Again, as a result slope selection is
achieved, which has already been discussed in great detail in \cite{af96b}.

%----------------------------------------------------------
\section{Detachment and desorption}       \label{detachment}
One might assume that other mechanisms could lead to a zero in the
slope dependent current. In the following we will carry out an
analogous calculation with an adatom--detachment rate from steps and
inclusion of desorption \cite{vp95}. One might assume that both
processes generate an effective downward current which can compensate
for the Ehrlich--Schwoebel effect.  To investigate whether they are
sufficient to obtain slope selection (and to simplify notation) we
exclude the incorporation mechanism. Thus, the distinction of the two
regions on a terrace is not necessary.

The desorption of diffusing adatoms is easily incorporated including a
term $-\rho(x)/\tau$ in the diffusion equation (\ref{diffusion})
\cite{vp95}. In order to include detachment from steps we have to
replace boundary condition (\ref{Bup}) by 
\begin{equation}
  -D \rho'(-\ell/2)  =  \gamma - \frac{D}{a} \rho(-\ell/2)
\end{equation}
where $\gamma$ stands for the detachment rate from steps. Accordingly,
the boundary condition at downward step has to be corrected and reads now
\begin{equation}
  -D \rho'(\ell/2)  =  
       \frac{D}{\ell_1} \rho (\ell/2) - \gamma \frac{a}{\ell_1}.
\end{equation}
The overall slope dependent current becomes
\begin{equation}
 J(\ell) = \frac{\left( \Delta - 1 \right )
   \left( \ell_1 - a\right) 
   \left( \frac{a \gamma}{\tau} - D F \right)}
 {(\ell_1+a) \sqrt{\frac{D}{\tau}} (\Delta+1) 
   + \frac{a \ell_1}{\tau}(\Delta-1) + D  (\Delta-1)}
\end{equation}
where 
\[ \Delta = \e^{2 \ell/\sqrt{D\tau}} \]
has been introduced. Note that $\Delta$ is always greater than one.

To discuss the qualitative behaviour it is sufficient to look at the
numerator of $J(\ell)$ (the denominator is always positive). The
first important result is that no slope 
selection is possible. Only for $\ell=0$ the current is zero (of
course, there is no upward and downward current as well).

Even though there is no slope selection, one can discuss whether growth
will proceed via layer--by--layer growth ($J(\ell) > 0$ for all $\ell$,
{\it i.e.} terraces tend to grow larger) or a growth instability is
present ($J(\ell) < 0$ for all $\ell$, {\it i.e.}~particles are
preferably incorporated at the upper steps).

In the well known limit of negligible detachment or desorption rates
\cite{ks95} ($\gamma \rightarrow 0$ or $\tau \rightarrow \infty$)
the Ehrlich--Schwoebel effect alone determines the sign of
$J(\ell)$. As expected, for positive step edge barriers ($\ell_1 > a$)
growth becomes instable whereas negative values of $E_S$ stabilize
layer--by-layer growth. 

If we assume a positive $E_S$ in the general case we obtain a
critical current
\begin{equation}
  F_C = \frac{a \gamma}{D \tau}
\end{equation}
where the current changes its sign.

If one expresses the diverse rates as used in the Solid--on--Solid
simulations and sets $a = 1$ one obtains
\begin{equation}
F_C = \nu_d \e^{-(E_D+E_{\mbox{\scriptsize bind}})/\kb T}.
\end{equation}
where the desorption rate $\nu_d \e^{-E_D/\kb T}$ has been
introduced. \ebind\ represents the typical binding energy of a
detaching adatom. In 2+1 dimensions this should be approximately
$\ebind \approx 2E_N$. Using the model of the previous section, $\nu_d
= \nu_0$, and $E_D=1.1\ev$ (parameters which are a reasonable guess
for CdTe(001), \cite{sk98}) one obtains a critical flux $F_C = 0.004$
ML/s. However, this crossover should not be observable in experiments,
since at such low external fluxes the step flow of the preexisting
steps will dominate the surface evolution.

%-----------------------------------------------------------
\section{Saturation profile with infinite step edge barrier} 
                                                 \label{saturation}
In this section we will calculate the saturation profile for the model
with infinite step edge barrier. The discussion of the closure of the
bottom terrace already showed that in this limit the calculations
become very simple. As for the bottom terrace the dynamics of higher
steps become independent of the above lying terrace. The steps
$x_i(t)$ evolve according to 
\begin{eqnarray}
\frac{\mbox{d} x_1}{\mbox{d} t}(t) & = & - F (x_1(t) + \ri) \nonumber\\
\frac{\mbox{d} x_2}{\mbox{d} t}(t) & = & - F (x_2(t)-x_1(t)) \nonumber\\
%\frac{\mbox{d} x_3}{\mbox{d} t}(t) & = & - F (x_3(t)-x_2(t)) \nonumber\\
 &  \vdots &  \nonumber\\
\frac{\mbox{d} x_i}{\mbox{d} t}(t) & = & - F (x_i(t)-x_{i-1}(t))\\
 & \vdots & \nonumber
\end{eqnarray}
Measuring the time in units of $1/F$ ({\it i.e.} setting $F=1$ in the
above equations) the time to grow one monolayer is equal to one. In
addition, to simplify notation we will measure all lengths in
units of $\ri$. 

The solution for the lowest terrace has been given in 
{\it eq.}~(\ref{xbottom}). The solution for all steps is 
\begin{equation}         \label{xgeneral}
x_n(t) = \sum_{i=1}^{n} \frac{t^{n-i}}{(n-i)!} 
                          (1+x_i(0)) \e^{-t}  - 1
\end{equation}
as can be easily verified. If we want to calculate the steady state
saturation profile we have to require 
\begin{equation}
x_{i+1}(1) = x_i(0).
\end{equation}
It should be stressed that this is the only assumption: the surface
morphology is a self-reproducing structure.
If the upper steps $x_{i+1}$ after deposition of one monolayer would
be greater than $x_i(0)$ this would result in a flattening of the
surface. Otherwise the slope would become steeper. Using the solution
for the bottom terrace (\ref{xbottom}) we obtain the initial value 
\begin{equation}
x_1(0) = \e - 1
\end{equation}
when we require that the bottom terrace will be closed at time $t=1$.

For the upper terraces {\it eq.}~(\ref{xgeneral}) yields a recursion
relation
\begin{equation} \label{recursion}
x_n(1) + 1 = x_{n-1}(0) + 1 = \sum_{i=1}^n \frac{1 + x_i(0)}{(n-i)! \e} 
\end{equation}
which can be solved as described in the appendix using the generating
function. As a result one obtains the initial positions of 
the steps on an infinite symmetric step profile. Every time the bottom
terrace is closed the steps (with new indices) are located at these
positions. 

\begin{table}
\begin{tabular}{r|r|c}         \label{tab_xi}
 i & $x_i(0)$ & $x_i(0)-x_{i-1}(0)$ \\
\hline
1 & $\e-1$ & 1.71828\\
2 & $\e^2-\e-1$ & 1.95249\\
3 & $\e^3-2\e^2+\frac{1}{2}\e -1$ & 1.99493\\
4 & $\e^4-3\e^3+2\e^2-\frac{1}{6}\e - 1$ & 2.00009\\
5 & $\e^5-4\e^4+\frac{9}{2}\e^3 - \frac{4}{3}\e^2 + \frac{1}{24}\e -1$
& 2.00007\\ 
\end{tabular}
\caption{ \label{tablepos}
  Analytical expressions of the saturation profile step
  positions and the numerical values of the terrace widths.}
\end{table}

In table \ref{tablepos} we show the analytical expressions for the
step positions derived from the generating function. In addition, the
numerical values of the terrace widths are shown. With growing index
the terrace widths are rapidly approaching two. Even though they
oscillate around this value it can be shown that 
\begin{equation}   \label{limestwo}
\lim_{i \rightarrow \infty} \left(x_i(0)-x_{i-1}(0)\,\right) = 2.
\end{equation}
Note that we measure the lengths in units of \ri. Hence we
predict a slope selection with slope $1/(2\ri)$ as derived from the
simple argument of section \ref{slopeselection}.

The derivation shows that the selected slope is controlled by the
closure of the bottom terrace only. Another length scale is the
nucleation length. It is defined by the typical length of a top
terrace at which nucleation of an island occurs \cite{vpt92}. This
length scale is responsible for the rounding of the towers in
fig.~\ref{profiles}b and causes a perturbation of the
steady--state saturation profile.

\section{Conclusion}
We have investigated the effect of an incorporation mechanism on the
morphology of growing surfaces. The inclusion of an incorporation
mechanism in a 1+1 dimensional  BCF--theory as well as in SOS
computer simulation in 2+1 dimensions is necessary in order to obtain
slope selection and a coarsening process. We were able to derive
analytically the temperature dependence of the selected slope. We
found that the Ehrlich--Schwoebel barrier alone controls the
temperature dependence. In the limit of an infinite step edge barrier
we derived the steady state saturation profile. In this case the
resulting mound morphology is controlled by the closure of the bottom
terrace.

\begin{appendix}
\section{Generating function}
To simplify notation we introduce the shifted step positions $b_j =
x_j(0) +1 $. We will try to extract the generating function 
\begin{equation}
  f(z) = \sum_{j=0}^\infty b_j  z^j
\end{equation}
for the shifted step positions. Clearly, the $b_j$ are only of
physical meaning if $j>0$ and $b_0$ can be chosen arbitrarily.
Starting from equation (\ref{recursion}) 
\begin{eqnarray}
b_{n-1} & = &  \sum_{i=1}^n \frac{b_i}{(n-i)!~\e} 
           \mbox{ for all $n \ge 2$}\\
\Rightarrow \; \;
  \e~z~b_{n-1}~z^{n-1} & = &  \sum_{i=1}^n \frac{b_i~z^n}{(n-i)!} \\
\Rightarrow \;
\e~z~\sum_{m=1}^\infty   b_{m}~z^{m} 
        & = &  \sum_{n=2}^{\infty} \sum_{i=1}^n \frac{b_i~z^n}{(n-i)!}
\end{eqnarray}
Choosing $b_0 = 0$ and using $b_1 = \e$ we arrive at
\begin{eqnarray}
\e~z~f(z) & = & 
  \sum_{n=0}^{\infty} \sum_{i=0}^n \frac{b_i~z^n}{(n-i)!} - \e~z\\
\Rightarrow \;
\e~z~f(z)  & = & f(z)  \e^z - \e~z
\end{eqnarray}
Thus, we finally obtain
\begin{equation}
 f(z) = \frac{z}{\e^{z-1} - z}.
\end{equation}
The lowest coefficients 
\begin{equation}
  b_j = \frac{1}{j!} \left. \frac{\partial^j f}{\partial z^j} \right|_{z=0}
\end{equation}
derived from the generating function are
shown in table (\ref{tab_xi}). In addition, the generating function
can be used to formally prove equation (\ref{limestwo}).

\end{appendix}

\begin{acknowledgement}
This work has been supported by
the Deutsche For\-schungs\-ge\-mein\-schaft DFG through SFB~410. 
\end{acknowledgement}

\bibliography{/users1/schinzer/Arbeit/Preprints,/users1/schinzer/Arbeit/Literatur,/users1/schinzer/Arbeit/NeueLit}
\bibliographystyle{unsrt}

\end{document}